\documentstyle{article}
\begin{document}
\renewcommand{\thefootnote}{\fnsymbol{footnote}}
\begin{center}
{\bf Some remarks on $N=2$ extended supersymmetric Yang-Mills theories 
and Seiberg-Witten duality} \\[3ex]
Rainald Flume\footnote[2]{Supported in part by the project of a Bonn/Wroclaw 
collaboration in the framework of the Polish/German university exchange 
program}\footnote[1]{Talk presented at the IX-th Max Born Symposium, 
held in Karpacz, Poland, September 1996; preprint Bonn-TH-97-04, February 
1997}\\
Physikalisches Institut\\
der Universit\"at Bonn\\
Nussallee 12\\
D-53115 Bonn\\
F.R.G.
\end{center}
The field content of a $N=2$ extended supersymmetric gauge theory comprises 
a complex scalar field $\phi$, a doublet of Weyl spinors $\chi_i , i = 1,2$ 
(and their chiral conjugates) a vector field $A_\mu$ and possibly an 
auxiliary field $X$ (to close the supersymmetry algebra off-shell). All fields 
are supposed to be multiplets transforming under the adjoint representation 
of the underlying gauge group, which will be taken as $SU(2)$ in the 
following. In the so-called Wess-Zumino (WZ) gauge the Lagrangean of the 
theory under consideration reads as
\begin{eqnarray}
\cal L &=& tr \left\{ - \frac{1}{4} F^2_{\mu \nu} + D_\mu \phi D^\mu \phi^* \right.
\nonumber\\
& + & i \bar{\lambda}_i \bar{\sigma}_\mu D^\mu \lambda^i + 
g \phi^* \left[ \chi _i , \chi^i \right] 
\nonumber\\
&+& \left. g \phi \left[ \bar{\chi}_i , \bar{\chi}^i \right] + g^2 
\left(\left[ \phi^* , \phi \right] \right)^2 + \frac{1}{2} X^2 \right\}
\end{eqnarray}
where $F_{\mu \nu}= \partial_\mu A_\nu - \partial_\nu A_\mu  + 
[A_\mu A_\nu ], g = $ gauge coupling \ldots etc. 

Here we make use of the standard notations of ref [1] .

The supersymmetric theory given by this Lagrangean ist first of all an 
ordinary gauge theory with peculiar couplings of the scalar and spinor 
fields. Supersymmetry is not an obvious invariance of $\cal L$. The selected 
WZ gauge has in fact a bad reputation what concerns this aspect. The 
reason is that supersymmetry is preserved in this gauge only modulo field 
dependent gauge transformations. The symmetry operations generate therewith 
an open algebra which seemed hopelessly difficult to control. (The choice  
of the WZ gauge is didicated by the technical advantages for instanton 
calculations to be discussed below). So-called supersymmetric gauges which 
are designed to cope with this problem imply the introduction of a bunch 
of subcanonical auxiliary fields. Those have their appearance in a rather 
complex non-polynomial (but manifestly supersymmetric) fashion. On the basis 
of manifest supersymmetry supergraph techniques have been developed [2], [3] 
allowing the derivation of the famous non-renormalization theorems of 
supersymmetric field theories. The specific non-renormalization theorem for 
$N=2$ theories has been established in [4].

A new approach to the renormalization of supersymmetric gauge theories in 
WZ gauges has been developed recently [5], [6], [7], [8]. The above mentioned 
problem of an open symmetry algebra has been overcome by the introduction 
of an extended BRS formalism, which combines the well known conventional 
Slavnov-Taylor (ST) operation of BRS with the generators of supersymmetry 
and space-time translations, the latter being parametrized by space-time 
independent ghost fields. The extended operator is by construction 
nilpotent (as the conventional BRS-ST operator). It opens the possibility 
to reduce the renormalization to an algebraic problem -- the determination 
of the cohomology of the extended ST operator. One may also nourish the 
hope that the non-renormalization theorems come into the range of the new 
method.

The formal basis of the particular $N=2$ non-renormalization theorems may be 
found in the fact that the Lagrangean (1) can be represented (up 
to total derivatives) 
as the fourfold chiral supersymmetry variation of an operator of 
canonical dimension two which for itself is invariant under anti-chiral 
transformations:
\begin{eqnarray}
\cal L &\simeq & \Delta^2 x , \bar{\delta} x = 0 \\
\Delta &=& \sum\limits_{i, \alpha} \delta^{i, \alpha} \delta_{i , \alpha}
\nonumber\\
x &=& tr \phi^2  \nonumber
\end{eqnarray}
where $\simeq$ denotes equality modulo total derivative terms. 
$\delta^{i , \alpha} (\bar{\delta})$ designates a chiral (anti-chiral) 
supersymmetry transformation. The spinor label $\alpha$ and the 
extendedness label $i$ are lowered (resp raised) by the two-dimensional 
skew symmetric tensor (cf. [1]). The Lagrangean can also be 
represented as the fourfold anti-chiral variation of the complex conjugate 
operator $x^*$, which is left invariant by chiral transformations:
\begin{eqnarray}
\cal L &\simeq& \bar{\Delta}^2 x^* , \delta x^* = 0\\
x^* &=& tr (\varphi^{*2} ) \nonumber\\
\bar{\Delta} &=& \bar{\delta} \cdot \bar{\delta}.\nonumber
\end{eqnarray}
Eq's (2) and (3) may be used to ''improve'' the ultraviolet behaviour of 
Green's functions by transferring susy variations with the help of 
Ward identities from interaction vertices to external fields. Exploiting 
this mechanism to the end one may extract four derivatives acting on 
external fields. (Each supersymmetric variation carries the canonical 
dimension $\frac{1}{2}$. One extracts altogether eight supersymetric
variations). Terms with a smaller number of derivatives $(< 4)$ will not 
appear in the effective low energy Lagrangean. The latter formal statement 
is the content of the non-renormalization theorem.

The argument for non-renormalization based on the use of supergraphs 
excludes one-loop contributions since at this level ''ghost-for-ghost'' 
terms emerge which are not covered by the argument making use of supergraph 
combinatorics. An equivalent reasoning for the breakdown of the above 
given formal derivation in one-loop approximation has to be worked 
out yet.
\vspace{1cm}

One may introduce, for the sake of a concise description of extended 
supersymmetry, a $N=2$ chiral superfield with two independent two-component 
Grassmann parameter $\theta _1 , \theta_2$  [9], which collects all 
fields figuring on the Lagrangean $\cal L$;
\[
\Psi = \phi + \theta_i \chi^i + \ldots  + \theta^2_1 \theta^2_2 
\left( 4 D_\mu D^\mu \phi - 2 \phi [\lambda_i , \lambda^i] + 
4 g^2 \left[\phi [\phi , \phi^*]\right] \right)
\]
The Lagrangean $\cal L$, Eq. (1), may be quoted as
\[
{\cal L} = Im \left( \frac{1}{2} \tau tr \Psi^2 \right)
\]
with 
\[
\tau = \frac{\vartheta}{2 \pi} + \frac{4 \pi i}{g^2}
\]
where $\vartheta$  denotes the topological vacuum angle and $g$ is the
 gauge coupling constant as before. The most general $N=2$ invariant local and 
hermitean expression in the fields enumerated at the beginning of this note 
can be represented as the imaginary part of an arbitrary ''holomorphic'' 
function $F(\Psi)$, the so-called $N=2$ prepotential. (Holomorphicity means 
here that $F$ depends on $\Psi$ but not on the chiral conjugate of $\Psi$.)

Seiberg and Witten (SW) [10] consider the $N=2$ theory, given by Eq. (1) in 
the broken phase assuming for the scalar field $\phi$ a non-vanishing 
vacuum expectation value $a$. The original $SU(2)$ invariance is broken 
down to $U(1)$. It is known since some time [11], that the classical 
prepotential is  modified through a 1-loop term. Higher loop perturbative 
contributions are absent. This follows from a non-renormalization theorem. 
The classical and 1-loop contributions add up to
\begin{equation}
F_{Pert}(a^2) = \frac{i}{2 \pi} a^2 \ell n \frac{a^2}{\Lambda^2}
\end{equation}
where $\Lambda$ denotes a cutoff momentum. (We restrict our attention in (4) 
to the vacuum expectation value of $F$. The full operator expression 
is recovered by substituting in (4) for the constant $a$ the superfield 
$\Psi$). Eq. (4) is exact to all orders in perturbation theory. It is 
certainly reliable for $|a| \gg |\Lambda |$ since this is the realm of 
asymptotic freedom predestinated for perturbation theory.

But it cannot be, as noted by SW, the full answer since it has 
not built in the positivity properties required for a consistent field theory.
E.g., the  imaginary part of the second derivative of $F$ with respect 
to $a$ should be positive as it is the square of an effective coupling 
constant. This is notoriously no longer the case if $a$ becomes of the 
order of $\Lambda$ in Eq. (4).

SW guess ingenously the correction to the perturbative result (4). Their 
proposal amounts to a function of which the second derivative is the well 
known elliptic modular function [12]. The motivation for this guess 
has been described in depth in the original paper -- in particular the 
dynamical realization of Maxwell-Dirac duality -- and reviewed many times 
since then.  So I confine myself to a few remarks.
\renewcommand{\labelenumi}{(\roman{enumi})}
\begin{enumerate}
\item The SW prepotential can be represented in the form of an expansion
\[
F(a)^2 = F_{pert} + a^2 \sum\limits_{k=1}c_k 
\left(\frac{\Lambda^2}{a^2} \right)^{2k}
\]
The coefficients $c_k$ are fixed through the SW conjecture. Seiberg argued 
on the other hand qualitatively [11] that instanton configurations should 
contribute in semiclassical approximation. The concrete 1-instanton 
[13]- [15] and 2-instanton calculations [15] -- ($k$-instanton configurations 
give rise to a contribution to $c_k$) -- lead to the identification of the 
SW-determined coefficients $c_1$ and $c_2$ and those extracted from the 
instanton calculation. The concrete instanton calculations are to be 
taken as a confirmation of the SW proposal. They moreover indicate that the 
$N=2$ prepotential is exactly saturated by instanton configurations.
\item I am tempted to speculate that the appearance of a ''nice'' function, 
the elliptic modular function, in the SW proposal is genuinely related to the 
quasiclassical exactness eluded to in the previous remark.
\item The term ''quasiclassical'' used in the previous remark is not 
entirely precise. The instanton configuration with associated scalar fields 
taken in [13] - [15] as starting points of the calculations are not 
saddle points of the $N=2$ action. To call them approximate saddle points 
is not
 appropriate either since the result of the calculation is supposed to be 
exact. It has to be noted in any case that the mechanism responsible for the 
restriction of the functional integral to instanton configurations 
has not been explained convincingly 
to date. Here comes the Duistermaat-Heckman (DH) integration formula [16] 
as a potential finite dimensional analogue to mind. (An account of the DG 
formula easily accessible for non-mathematicians may be found in [17]). I 
conclude with a sketch of the DH formula.

Let $M$ be a compact $2n$-dimensional manifold without boundary. $M$ is 
supposed to have a symplectic two-form $\omega$ (i.e., $\omega$ is closed 
and non-degenerate). DH propose the evaluation of integrals of the form 
\[
J(H) = \int\limits_M \frac{\omega^n}{n!} \exp (H),
\]
where $H$, the ''Hamiltonian'', is a sufficiently smooth function. To proceed 
along the lines of DH I recall the notion of a Hamiltonian vector field 
denoted $rot H$, which is defined by
\begin{equation}
dH(x) = \omega (x, rot H)
\end{equation}
where $x$ is an arbitrary vector field on $M$. One can find an exact one-form 
$d \varphi$ (dual to $rot H$). s.t. at points of non-vanishing $rot H$ 
\begin{eqnarray}
\frac{\omega^n}{n!} &=& d \varphi \wedge \omega_{rot H} \wedge 
\frac{\omega^{n-1}}{(n-1)!} \nonumber\\
&=& dH \wedge d \varphi \wedge \frac{\omega^{n-1}}{(n-1)!}
\end{eqnarray}
holds, where the notation 
\[
\omega_\Psi (x) \equiv \omega (\Psi , x)
\]
is used. Eq. (5) has been used for the second equality in (6). Let me now 
assume that the critical points of $H$, that is, the points of vanishing 
$rot H$, are concentrated on a low-dimensional submanifold $X_{crit}$ of 
$M$. Denoting with $M_\epsilon$ the subvariety of $M$ with a distance 
larger than $\epsilon$ from $X_{crit}$ one obtains with Stokes theorem and 
Eq. (6)
\begin{equation}
J(H) = \int\limits_M e^H \frac{\omega^n}{n!} = \lim\limits_{\epsilon \to 0} 
\int\limits_{M_\epsilon} e^H \frac{\omega^n}{n!} = 
\lim\limits_{\epsilon \to 0} \int\limits_{\delta M_\epsilon} 
e^H d \varphi \wedge \frac{\omega^{n-1}}{(n-1)!} ,
\end{equation}
with $\delta M_\epsilon$ denoting the boundary of $M_\epsilon$. The 
last equality means that the integral becomes localised at the submanifold 
of critical points of $rot H$. One can rather easily evaluate the r.h.s of 
Eq. (4) for the case that $X_{crit}$ consists of isolated points. The 
result is then formally identical with a quasiclassical approximation of 
$J(H)$ with the isolated critical points handled ''as if'' they were 
stationary points of $H$. Note however, that criticality is not synomous 
with stationarity.
\end{enumerate}
\noindent
{\bf Acknowledgement}\\
I am indebted to L. O'Raifeartaigh and I. Sachs for an initiation to 
Maxwell-Dirac duality and I thank K. Sibold fo a hint to references [5] - 
[8].\\[2ex]
\noindent
{\bf References}\\
\renewcommand{\labelenumi}{[\arabic{enumi}]}
\begin{enumerate}
\item J. Wess and J. Bagger, Supersymmetry and Supergravity, Princeton Series 
in Physics (1983).
\item M. T. Grisaru, W. Siegel, and M. Rocek, Nucl. Phys. {\bf B159} 
(1979) 429.
\item M.T. Grisaru and W. Siegel, Nucl. Phys. {\bf B201} (1982) 292.
\item P.S. Howe, K.S. Stelle, and P.K. Townsend, Nucl. Phys. {\bf B236} 
(1984) 125. 
\item P.L. White, Class. Quantum Grav. {\bf 9} (1992) 1663.
\item N. Maggiore, Off-shell formulation of $N=2$ Super Yang-Mills theories 
coupled to matter without auxiliary fields, Rep-th/9412092.
\item N. Maggiore, Algebraic renormalization of $N=2$ Super Yang-Mills 
theories coupled to matter, hep-th/9501057.
\item N. Maggiore, P. Piguet, and S. Wolf, Algebraic Renormalisation of $N=1$ 
Supersymmetric Gauge Theories, hep-th 9507045.
\item R. Grimm, M. Sohnius, and J. Wess, Nucl. Phys. {\bf B133} (1978) 275.
\item N. Seiberg and E. Witten, Nucl. Phys. {\bf B426} (1994) 19, 
{\bf B430} (1994) 485.
\item N. Seiberg, Phys. Lett. {\bf B206} (1988) 75.
\item Erdelyi, Higher Transcendental functions, Vol 3, McGraw Hill (1955).
\item D. Finnell and P. Pouliot, Nucl. Phys. {\bf B453} (1995) 225.
\item K. Ito and N. Sasakura, Phys. Lett. {\bf B382} (1996) 95.
\item N. Dorey, V.V. Khoze, and M.P. Mattis, Phys. Rev. {\bf D54} 
(1996) 2921.
\item J.J. Duistermaat and G.J. Heckman, Invent. Math. {\bf 69} (1982) 259.
\item R.F. Picken, J. Math. Phys. {\bf 31} (1990) 616.
\end{enumerate}
\end{document}